# Effect of Pore Formation on Redox-Driven Phase Transformation

Xuyang Zhou[1,*], Yang Bai[1,*], Ayman A. El-Zoka[1], Se-Ho Kim[1], Yan Ma[1], Christian H. Liebscher[1], Baptiste Gault[1,2], Jaber R. Mianroodi[1], Gerhard Dehm[1], Dierk Raabe[1,*]

[1] Max-Planck-Institut für Eisenforschung GmbH, 40237 Düsseldorf, Germany.

[2] Department of Materials, Royal School of Mines, Imperial College London, SW7 2AZ London, UK.

*Correspondence to: x.zhou@mpie.de; y.bai@mpie.de; d.raabe@mpie.de

**Abstract:**

When solid-state redox-driven phase transformations are associated with mass loss, vacancies are produced that develop into pores. These pores can influence the kinetics of certain redox and phase transformation steps. We investigated the structural and chemical mechanisms in and at pores in a combined experimental-theoretical study, using the reduction of iron oxide by hydrogen as a model system. The redox product (water) accumulates inside the pores and shifts the local equilibrium at the already reduced material back towards re-oxidation into cubic-$Fe_{1-x}$O (where x refers to Fe deficiency, space group $Fm\bar{3}m$). This effect helps to understand the sluggish reduction of cubic-$Fe_{1-x}$O by hydrogen, a key process for future sustainable steelmaking.

Redox-driven phase transformations occur in crystalline materials when a parent phase becomes metastable due to the chemical driving force imposed by a redox process [1]. In solid state redox processes such transformations are usually accompanied by mass gain (e.g. in corrosion) or mass loss (e.g. in oxide reduction) of the solid parent phase. Here we discuss the latter scenario, i.e. oxygen loss during reduction of a solid oxide. Changes in temperature, pressure, chemical composition, structure, or topology can strongly influence the kinetics of such reduction processes and the associated phase transformation steps [2-5]. Recent work [6,7] has shown that reduction kinetics depend not only on transport coefficients and nucleation barriers in ideal, defect-free oxides, but are also significantly influenced by microstructural features that are present in the parent phase or evolve in it during reduction. The propagation of the phase boundary between oxidized and reduced domains is highly dependent on the defect density [8-10]. In particular, it has been shown that dislocations, interfaces, and pores can lower critical nucleation energies and result in order-of-magnitude higher diffusion coefficients as reaction species move along the defects at higher rates than in perfect bulk lattices [8,9]. While the role of interfaces has been discussed in the literature [6,8,9,11], the influence of pores has not yet been investigated at the level of a single defect.

We chose the direct reduction of iron oxide with $H_2$ as model system to study such a redox-driven solid state phase transformation since it is the precondition and key process behind a future sustainable global steel industry [6,11-13]. H-based direct reduction is a fossil-free approach to iron production in which solid iron oxides are exposed to gaseous $H_2$. Iron oxide is today reduced by CO for the production of roughly 1.8 billion tons of steel per year, causing approximately 7% of the total global $CO_2$ emissions. To mitigate this huge greenhouse gas burden, the use of sustainably produced $H_2$ as reductant is re-emerging as an alternative and carbon-neutral synthesis pathway [12]. Because of the gigantic annual steel production, $H_2$ utilization strategies that yield high metallization at fast reduction kinetics are urgently required, to make such processes commercially viable and efficient. A key to understanding the rate-limiting solid-state transport and reaction mechanisms behind this approach is the study of the role of the microstructure and specifically of the pores that form during the reduction process due to the mass loss of oxygen.

In partially reduced iron oxide, there are two types of pores, namely, (a) those formed before the reduction during the sintering of the fine oxides into commercial pellets and (b) those formed during reduction, due to the mass and volume changes, associated with the transformations from trigonal-$Fe_2O_3$ (space group $R\bar{3}c$) to cubic-$Fe_3O_4$ ($Fd\bar{3}m$) then to cubic-$Fe_{1-x}O$ (where x refers Fe deficiency, $Fm\bar{3}m$) and finally to body-centered cubic (BCC) – Fe ($Im\bar{3}m$). To study the structure and chemistry of the latter types of pores as they evolve during reduction at a single-defect and nanoscopic scale, we used single-crystalline cubic-$Fe_{1-x}O$ as an otherwise defect-free well-defined reference material, to avoid pores and other defects inherited from sintering.

The aim of our study is twofold: firstly, we study the structural, crystallographic, size distribution, and topological features of pores and the structures surrounding them at different positions relative to the free surface to better understand their role in the overall H-based reduction. Secondly, we exploit individual pores as confined nano-labs with locally well-defined chemical and structural boundary conditions to better understand the role of water and re-oxidation on the sluggish transformation from cubic-$Fe_{1-x}O$ to BCC-Fe. Note that this last stage in the transformation sequence is up to a factor of ten slower than the preceding phase transformations, e.g. the reduction of trigonal-$Fe_2O_3$ to cubic-$Fe_3O_4$ or cubic-$Fe_3O_4$ reduction to cubic-$Fe_{1-x}O$. The total achievable metallization (i.e. the degree of reduction) does mostly not exceed 90-95% [8].

To achieve this goal, the state-of-the-art four-dimensional scanning transmission electron microscopy (4D-STEM) method was used to study the phase state and structure near the pore surface, at a spatial resolution of about 1-2 nm [14-16]. Such fine crystallographic information is otherwise difficult to obtain by conventional methods such as selected area diffraction. Compared to existing phenomenological studies in the literature [17-19], we focus here on explaining some of the underlying nanoscale physics mechanisms responsible for the sluggish reduction commonly observed at temperatures between 600ºC and 800ºC during H-based direct reduction of iron oxides.

In this work, we found a pore size gradient, ranging from tens of nanometers to a few micrometers, developed during reduction through the thickness of the single-crystalline cubic-$Fe_{1-x}O$ with increasing distance from the free surface. On the inner rims of the pores, cubic-$Fe_{1-x}O$ has formed, although all the adjacent regions had been already fully reduced into pure BCC-Fe by the hydrogen. A rough estimation of the total affected volume fraction for this re-oxidized cubic-$Fe_{1-x}O$ phase surrounding the pores at their inner surfaces gives 5%, which is significant when considering the huge absolute numbers in the field of steelmaking. The experimental observation provides relevant information about the reduction (and also about the undesired re-oxidation) processes, confined in the pore region. With the boundary condition determined from the experiments, we performed phase-field simulations and found that the formation of a cubic-$Fe_{1-x}O$ phase on the inner surface of the isolated pores is caused by re-oxidation of the previously reduced material through the water that had accumulated inside of the pores. The knowledge gained from this study could guide the development of oxide feedstock and reactor strategies with regards to topological pore, size and percolation features suited for higher reduction kinetics and metallization in the field of H-based green steelmaking.

We characterized the structure and composition of the as-received sample using high-resolution scanning transmission electron microscopy (HR-STEM) and electron energy loss spectroscopy (EELS, presented in the supplemental material). We confirmed that the initial state of the sample is single-crystalline cubic-$Fe_{1-x}O$, see Fig. S1. This cubic-$Fe_{1-x}O$ was reduced at 700°C for 2 hours under a $H_2$ flux of 30 L/h. The heat-treatment profile is shown in Fig. S2. The microstructure of the reduced sample has been mapped in the scanning electron microscope (SEM) (Fig. 1), where the dark contrast indicates the pores formed during H-based reduction. The top view of the reduced sample shows the formation of pits with a size of several micrometers on the sample surface (Fig. 1a). The cross-sectional view (Fig. 1b) reveals a visual gradient of pore size ranging from nanometers to tens of micrometers across the thickness of the reduced cubic-$Fe_{1-x}O$. The regions near the top surface of the reduced sample appear relatively dense. John and Hayes [13] reported that the formation of a dense Fe layer substantially reduces the reduction rate as the evading oxygen must diffuse through it.

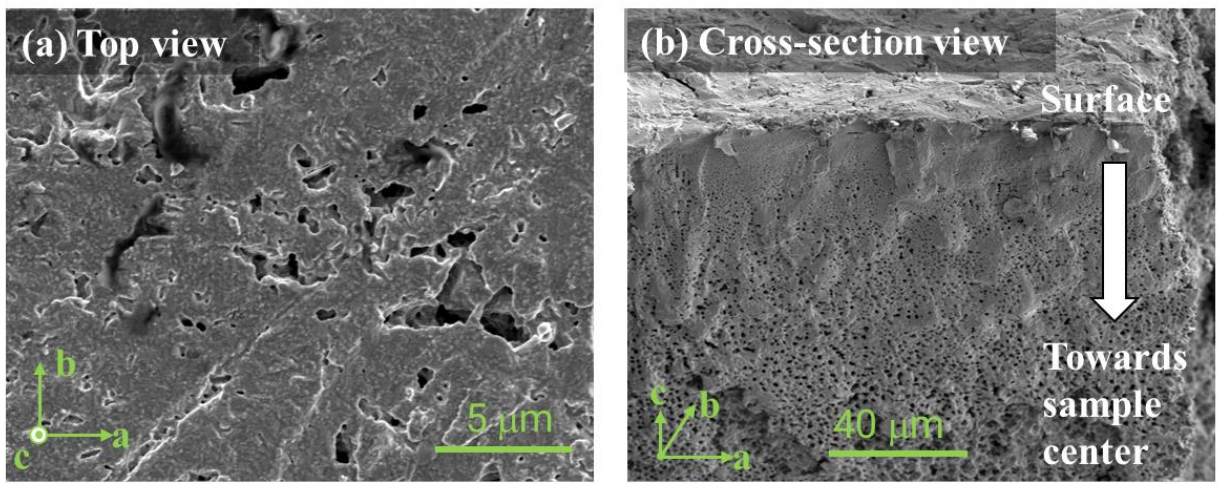

**Figure 1** Microstructure of a reduced single-crystalline cubic-$Fe_{1-x}O$ imaged by using SEM after partial reduction at 700°C for 2 hours under a $H_2$ flux of 30 L/h **(a)** Top view and **(b)** Cross-section view. The surface exposed to the reducing $H_2$ environment is perpendicular to the c direction.

We then employed (scanning) transmission electron microscopy (S/TEM) for the nanoscale analysis. Fig. 2a shows the high-angle annular dark-field (HAADF) STEM image of the focused ion beam (FIB) thinned TEM lamella lifted out from the top surface of the reduced sample. The HAADF contrast is primarily dependent on the atomic number Z and sample thickness. Although we chose a small camera length (12 cm), slight grain-to-grain contrast differences can still be seen due to the contribution of diffraction contrast. Nevertheless, the strong image contrast in Fig. 2a is predominantly caused by a change in sample thickness. The dark regions represent empty areas *i.e.,* pores. Areas of dark gray contrast are due to small pores in the Fe matrix. We found that some small pores assumed a facetted Wulff shape [20], see the regions marked with red arrows in Fig. 2a. In addition, some unwanted curtaining of the FIB lamella is present. Nevertheless, an increasing gradient in pore size was clearly observed from the sample surface to the interior, similar to our SEM observation (Fig. 1b), but with a much smaller pore radius of approximately several tens of nanometers. In addition to these distinct pores, there are also gray areas that appear to form channels between adjacent pores, which have been marked by blue arrows in Fig. 2a.

The 3D view of the pores is shown in Fig. 2b, which is an electron tomography (ET) reconstructed from a tilt series of HAADF images (-70° to 70° with 5° tilt increments) [21]. The resolution of the 3D imaging is determined by the number of tilt images and the particle size, in this case the pore size, according to the Crowther criterion [22]. Therefore, we could only image pores with a size above ten nanometers. The regions in royal blue color are the Fe matrix, while the salmon-colored ones are the pores. Figs. 2c&d show the magnified HAADF image and ET reconstruction of the region outlined in blue in Fig. 2a. To illustrate the 3D reconstructions, we have rendered two videos (supplemental material Video_1 and Video_2). Several findings emerge from the HAADF and ET reconstructions: (1) There are isolated small pores with a size of a few tens of nanometers distributed both inside the Fe grains and at the grain boundaries. (2) The pores at the grain boundaries or triple junctions appear larger than the pores in the grain interior. (3) These small pores increase in size from the sample surface toward the interior. (4) It can be seen that some large pores are interconnected and form an open percolation structure, which is a precursor to the interconnected pores that form the continuous network at the surface. The latter feature is particularly important for green steel production as connected pores allow for outbound diffusion of the water while close pores do not.

We then qualitatively compared the local composition using energy dispersion X-ray spectroscopy (EDS). Fig. 2e shows the intensity maps of the Fe-K shell and the O-K shell. Two large dark areas appear in the upper part of the two intensity maps and correspond to the pores. There are also some small dark areas in the Fe-K intensity map, but they have higher intensity in the O-K map, indicating high O content, showing that oxidation occurred near the pore.

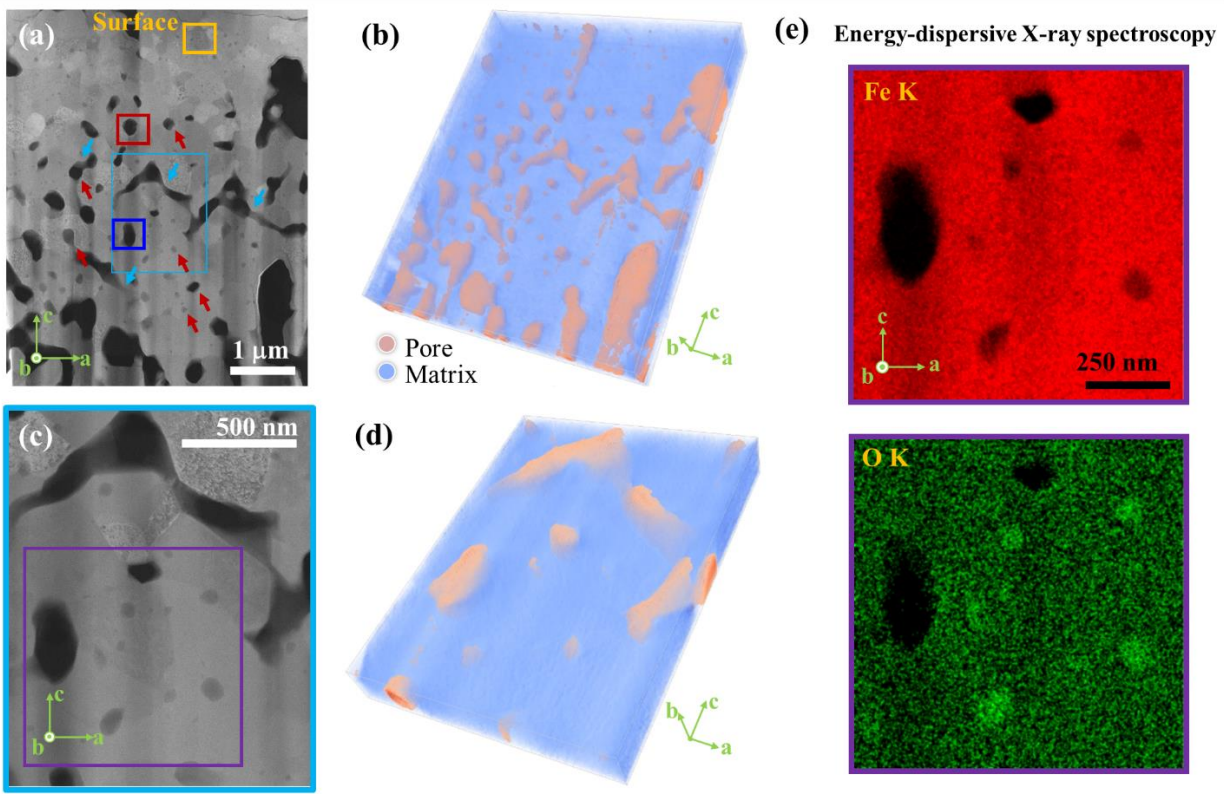

**Figure 2 (a)** A representative HAADF image of the H-reduced cubic-Fe$_{1-x}$O sample. Areas marked with red arrows appear to assume a facetted Wulff shape. Regions of interest for 4D-STEM scans (Fig. 3) are outlined in color. The rectangular areas in orange, red, and blue are regions close to the surface, in the middle of the lamella, and away from the surface, respectively. Some channels between adjacent pores are marked by blue arrows. **(b)** 3D reconstruction of the ET results. The regions in royal blue color coding are the BCC-Fe matrix, while the salmon-colored ones are the pores. The reconstruction is based on 29 HAADF images (tilting from -70° to 70° for every 5°). **(c & d)** are magnified images of the rectangular area highlighted in blue in a. **(e)** EDS mapping of the purple rectangular region in c.

The local structure near the pore areas has been measured by 4D-STEM [14-16]. Fig. 3a shows the phase and orientation mapping near three pore regions (1) close to the surface (in the orange boxes, i, iv and vii), (2) in the middle of the lamella (red boxes, ii, v, and viii), and (3) away from the surface (blue boxes, iii, vi, and ix). We first present the virtual bright-field images reconstructed from the intensities of the transmitted beam in each diffraction pattern in Fig.3a i-iii. In the phase maps (Fig. 3a iv-vi), the cubic-$Fe_{1-x}O$ phase (color-coded in green) has been found mainly located at the pore surfaces and at the triple junctions. The orientation maps in Fig. 3a vii-ix show that the originally single-crystalline cubic-$Fe_{1-x}O$ changed into a polycrystalline phase with a grain size of about 100 nm after H-based reduction. Fig. 3b contains selected diffraction patterns with the overlaid templates from either a BCC-Fe phase (# <u>1</u>, <u>3</u>, and <u>5</u>) or a cubic-$Fe_{1-x}O$ phase (# <u>2</u>, <u>4</u>, and <u>6</u>), the locations of which are highlighted in magenta color in Fig. 3a i-iii.

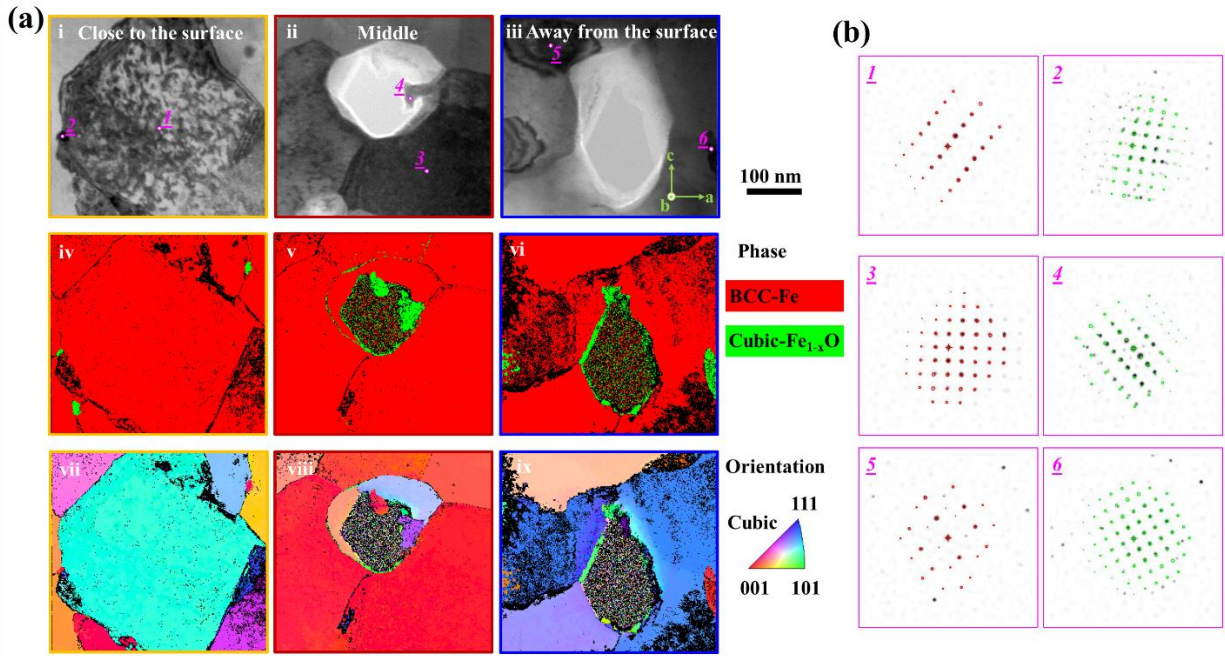

**Figure 3 (A)** 4D-STEM reconstruction to show **(i-iii)** the virtual bright-field images **(iv-vi)** the phase maps and **(vii-ix)** the orientation maps (i.e. inverse pole figure maps) of three regions outlined with colored frames in Figure 2a. In the phase and orientation maps, the regions with a confidence index below 0.1 are shown in black. **(B)** Selected diffraction patterns from points #<u>1</u>-<u>6</u> are highlighted in magenta in the virtual bright-field images (i-iii). The red and green spot matrices correspond to BCC-Fe and cubic-$Fe_{1-x}O$, respectively.

Studies have shown that the $H_2/H_2O$ ratio determines the morphology of the pores formed during the reduction [11,13]. However, how the local gas environment influences the reduction process and even leads to confined re-oxidation of the pore's inner rim is unknown. We therefore conducted phase-field simulations [7], where (a) the thermodynamics of the phase transition is captured by the underlying Landau energy form fitted to the Gibbs free energy curves for BCC-Fe and cubic-$Fe_{1-x}O$; (b) the kinetics by the elemental mobility in the different phases and the redox reaction rates; and (c) the boundary and initial conditions by real topologies mapped in the experiments, Fig. 4. Fig. 4a-i shows the simulated molar fraction of oxygen at different times during the reduction. As hydrogen diffuses into the sample, the transformation starts at the solid-gas interfaces. Therefore, the region close to the open pores experiences a lower oxygen concentration compared with the region far from the open channels. This is confirmed by Fig. 4a-ii, where the isolated pores are filled with about 15% water (for the details please see the supplemental material) due to the reduction reaction. However, in the open channels, water is transported out and replaced by hydrogen. Therefore, these channels almost always have lower water concentration compared to the closed pores. In contrast to this, once the water in the closed pores reaches the equilibrium value the reaction slows down, which further reduces the overall reduction rate as shown in Fig. 4a-iii&iv. As seen in this figure, in the regions near the closed pores, the phase transformation from cubic-$Fe_{1-x}O$ to BCC-Fe is slow. This is also seen in Fig. 4c, where the pressure change for different pores is plotted. Once the water fraction reaches the equilibrium value, the pressure profile reaches a plateau during the reduction process. The results also show that during the reduction reaction, the water vapor can reach about 4~5 atmospheres (~0.5 MPa) of pressure in the closed pores. It should be emphasized that this pressure is fairly low compared to the stresses induced by the phase transformation (around 1 GPa) [7]. Since the fracture release energy of steel is around hundreds of Joule per square meter, the crack propagation during the reduction and oxidation is therefore attributed to the phase transformation, and not to the pressure caused by the trapped water.

Once the sample is fully reduced, we enable the re-oxidation reaction in the simulations during the subsequent cooling process from 700°C to room temperature, based on the reaction model described in the supplemental material with the same boundary conditions as the reduction process. As we can see in Fig. 4b-i&ii, once the oxidation reaction starts, the trapped water in the closed pores will be consumed and a rim of oxide forms. Pores with different sizes experience different levels of water vapor. For instance, the amount of water vapor stored in pores-1, -2, and -3 is slightly different. As the re-oxidation proceeds during the simulated cooling, more water is consumed as shown in Fig. 4b-ii at t=2.0s and t=5.0s. This is also seen in Fig. 4b-iii, where a thin layer of cubic-$Fe_{1-x}O$ is generated around the closed pores.

The thickness of the circumferential cubic-$Fe_{1-x}O$ layer increases more during oxidation, until the water is completely consumed by the re-oxidation process. However, the thickness of a pore's individual $Fe_{1-x}O$ layer depends on the local re-oxidation kinetics. To further investigate the oxidation process, we plot the molar fraction of water vapor and also the phase fractions of cubic-$Fe_{1-x}O$ for the pores at different locations in Fig. 4d. Initially, due to the equilibrium, different pores experience the same fraction of water vapor, as shown in Fig. 4d by the different markers. As the re-oxidation reaction progresses from 700°C to 25°C, the closed pores with different sizes can lead to different cubic-$Fe_{1-x}O$ phase fractions around the pores, as shown in Fig 4d by the solid lines with different colors. This further indicates that the local water vapor and the size of the pore have a substantial impact on the re-oxidation process. Moreover, the pressure change of those pores is also slightly different, see Fig. S3.

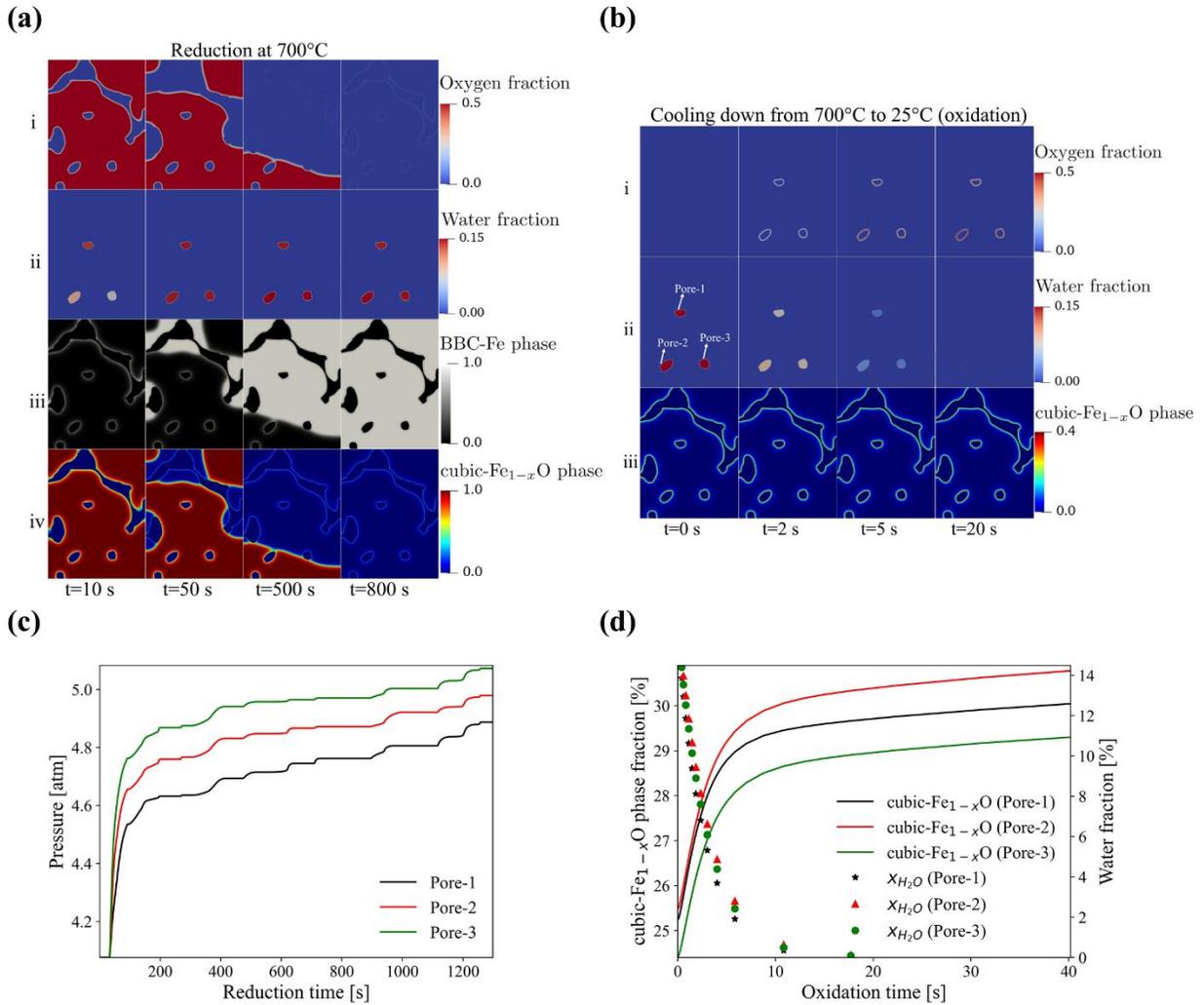

**Figure 4 (a) i-iv** top-to-bottom rows show the phase field simulation results for the molar fraction of oxygen and water, and the BCC-Fe phase and cubic-$Fe_{1-x}$O phase fractions, respectively, at different reduction time at 700°C. **(b) i-iii** Contour plots of the oxygen and water content, and cubic-$Fe_{1-x}$O phase evolution at different oxidation times during the cooling process from 700°C to 25°C. Since the cooling time is comparatively small, the oxidation reaction rate is treated as constant. **(c)** Pressure change of different pores under the reduction reaction at 700°C. **(d)** Fractions of water vapor and of the cubic-$Fe_{1-x}$O phase fraction of different pores during the simulated cooling, the initial number of the cubic-$Fe_{1-x}$O is based on the final status of the reduction process, and the initial water fraction is around 15% which is the equilibrium one in reduction.

In summary, we have studied the formation of pores during H-based reduction and their thermodynamic, structural, chemical, and kinetic roles in the redox-driven phase transformation of cubic-$Fe_{1-x}$O to BCC-Fe. The combined use of nanoscale microscopy and phase-field simulation revealed a new mechanism that might help to explain their rate-limiting effect on the reduction process: we found that the re-oxidation of BCC-Fe by the water that is trapped in isolated pores can lead to re-oxidation, i.e. through the formation of a cubic-$Fe_{1-x}$O phase at the pore surface, an effect which reduces the overall metallization and transformation rate. The re-oxidation effect is reduced if the pores have a percolating connection with each other and with free surfaces towards which the redox product water could escape. Based on this, we propose that creating a percolating pore (or micro-fracture) structure in feedstock materials used for H-based reduction of iron ores for green steel production can effectively increase reduction kinetics and improve metallization. Manipulating the percolation and connectivity of the gradually forming pore network during H-based reduction, due to the loss of oxygen, e.g., by changing the oxide's fracture toughness, grain size, interface cohesion and agglomerate size, could be practical solutions to accelerating the sluggish reduction kinetics, through outbound diffusion of the $H_2O$ instead of trapping it inside closed pores.

**Acknowledgements**

X.Z. acknowledges the support from the Alexander von Humboldt Foundation. X.Z. and C.H.L acknowledge funding by the German Research Foundation (DFG) through the project HE 7225/11-1. D.R acknowledges funding by the European Research Council (ERC) via the ERC advanced grant 101054368. Views and opinions expressed are however those of the authors only and do not necessarily reflect those of the European Union the ERC. Neither the European Union nor the granting authority can be held responsible for them.